# Light Sailboats: Laser driven autonomous microrobots

Anrdás Búzás[1], Lóránd Kelemen[1], Anna Mathesz[1], László Oroszi[1], Gaszton Vizsnyiczai[1], Tamás Vicsek[2], Pál Ormos[1]

[1]: Institute of Biophysics, Biological Research Centre, Hungarian Academy of Sciences, Temesvári krt 62. H-6726 Szeged, Hungary

[2]: Department of Biological Physics, Eötvös University, and

Statistical and Biological Physics Research Group of the Hungarian Academy of Sciences, Pázmány P. stny. 1A, H-1117 Budapest, Hungary

**Abstract**

We introduce a system of light driven microscopic autonomous moving particles that move on a flat surface. The design is simple, yet effective: Micrometer sized objects with wedge shape are produced by photopolymerization, they are covered with a reflective surface. When the area of motion is illuminated perpendicularly from above, the light is deflected to the side by the wedge shaped objects, in the direction determined by the position and orientation of the particles. The momentum change during reflection provides the driving force for an effectively autonomous motion. The system is an efficient tool to study self propelled microscopic robots.

There is great interest in self propelled microscopic robots, originating from both technology and fundamental science. Intensive development is directed towards technological applications primarily in the biomedical field, for tasks like material transport, local diagnostics, etc. Individual or collective motion of active biological systems (from bacteria to large animals) is also the target of a wide range of basic science studies due to remarkable dynamical phenomena in such non-equilibrium systems (for a recent review see Ref.1). The experimental studies require appropriate functional moving models. Different systems have been developed, depending on the requirements and the capabilities. The basic examples are of course the moving organisms themselves, they represent the various essential features as well: the mechanism of locomotion is contained in every unit, the energy for propulsion is provided in the medium and it is gained by the moving particles upon demand. Accordingly, in many experiments living bacteria are used (see, e.g., Refs. 2). However, the properties of the living objects are largely fixed and it is not possible to modify them significantly. In the process of testing specific theories there is also need for objects with different and controllable properties: size, shape, velocity, interaction between particles, etc. Consequently, there is need for artificial systems the properties of which can be arbitrarily determined and controlled. Various kinds of self propelled microscopic swimmers have been introduced, with most different propulsion means: chemical (6-8), light induced chemical (9), light induced thermal (10), electromagnetic (11). However, up till now the microscopic models have produced only limited results.

In this field in many cases size is not a principal requirement, after all the space where the particles are moving may not be limited. In many experimental approaches independent, autonomous robots are used, primarily in a two dimensional environment: robots rolling on a solid surface (12) or ships floating on a water surface (13), or even helicopters flying in 3 dimensions (14, 15). They have the obvious advantage that their properties can be defined in a practically arbitrary way, even quite complex behavioral pattern (propulsion, control, interaction) can be realized. These latter systems, however, are fairly complicated, the experimental area is also large. There is still need for possibly simple microscopic swimmers with efficient propulsion and well defined properties.

Here we introduce a simple yet very effective self propelled swimmer. The energy source for motion is light. The system is two dimensional, the particles are moving on a planar surface.

There are several properties the swimmers should have. The requirement concerning the propulsion is that it should i) result in a motion of the units that is regular to some extent, ii) in a direction corresponding to a "forward" motion of the object (there should exist a forward direction, fixed to the body of the swimmer), with possibly long persistence length. The energy supply for the motion should be continuous. The shape of the swimmer should be possible to vary, but in general an elongated shape is preferred.

Light has proven to be a practical means for the transport of microscopic particles. Optical micromanipulation has developed into an extremely powerful tool in biology (16). However, in the procedures demonstrated so far, either light pressure was used to transport particles (in this case objects are moving in the direction of the light propagation (17), not determined by their own orientation) or optical traps were used to grab and move particles in a direction determined by the motion of the trap itself (18). Such objects cannot be regarded as autonomous particles.

There is a still not reported possibility to harness the light energy to drive the microscopic particles in a way that fulfills the requirement for the motion of self propelled swimmers.

In the approach introduced in this work we use reflective wedge shaped particles, sliding on a flat horizontal surface (2D system). The area of motion is illuminated from above by a collimated homogeneous light beam, from a direction perpendicular to the surface of motion (see Fig.1.). The light is reflected on the surface of the wedge in a direction determined by the shape and position of each individual particle. The structure has a mirror symmetry and when moving it rests on one plane of the wedge while light coming from above is reflected to the side from the other plane transferring the momentum to the body. The momentum change during reflection provides a force sufficient to propel the particles, and a motion results in a direction determined by the object itself. Thus, effectively autonomously moving particles are obtained. The energy for motion is supplied by the light, it is available everywhere in the region of interest, and the particles use this energy in such a way that they move in the direction "they want" (as determined by their shape and position).

The used structures had several characteristic features necessary for efficient function. The angle between the sides of the wedges is 45 degrees to provide optimal momentum transfer. In addition, the structure has an elongated shape as seen from above – this is needed for the determination of the object orientation. Last, the aspect ratio of the body is such that when dropping to the surface it falls on the planar side of the wedge.

The scheme of the experimental system is sketched in Fig.1. The swimmers harvest the momentum of light analogous to the way sailboats use wind, we therefore call them light sailboats.

Fig.2.a. shows the exact geometry of the structures built and used in the experiments. We produced micrometer sized swimmers by two photon excitation photopolymerization (19, 20). A negative photoresist SU8 (Michrochem, Newton, MA, USA) was used as building material, it has excellent mechanical qualities after

polymerization. The polymerizing light was provided by a femtosecond fiber laser (C-Fiber A 780, Menlo Systems, Germany) operating at 100MHz frequency and 100 fs pulse length at 780 nm; the laser intensity used for the polymerization was about 5 mW at the back aperture of the focusing objective. We applied direct laser beam scanning for the photopolymerization. The polymerizing light was introduced into a Zeiss Axiovert35 microscope through the epifluorescence port. The laser focus scanning to draw the desired object was achieved by moving the sample along the pre-programmed path. The scanning unit was a combination of a 24-446 motorized table with an LSTEP 13 driver unit (Märzhäuser Wetzlar GmbH, Germany) for coarse movement and fine positioning was achieved with a 3D piezo scanner system (Physik Instrumente, Germany) consisting of a 100x100 μm travel length NanoPositioner stage and a 80μm travel length PIFOC microscope objective displacer. The photopolymerization setup was controlled by a computer running Labview software. With these components the resolution was about 100 nm in the x, y direction, and 500 nm in the z direction.

Fig.2. depicts the fabricated wedge structures. The structures were polymerized in a position such that the symmetry plane was perpendicular to the surface. It is advantageous for further processing that following polymerization the structures remain fixed to the glass surface. This was achieved by polymerizing two extra support rods of about 1 μm length between the structures and the glass surface. The support rods can be seen in Fig.2.c. After photopolymerization the surface of the wedges was made reflective by covering them with an Au metal mirror layer. The metal layers were deposited on the surface by sputtering (Emitech K975X, UK). In order to obtain a stable gold surface, two metal layers were applied. A 5 nm thick Cr layer was applied directly on the SU8 surface, this was covered by a 200 nm Au layer. Due to the position of the wedges on the carrier surface both planar sides received identical metal coverage. The resultant reflectivity is practically 100 % at the desired wavelength (1070 nm). Fig. 2.b shows the photopolymerized structures on the glass surface on which they were produced, already carrying the reflective gold surface cover. Finally, the wedges were removed from the carrier surface and introduced to the sample (Fig. 2.c. d.).

The motility experiments were performed in water environment. The motion was tested positively on planar glass surface (microscope slide), but substantial and irregular contact friction resulted in relatively slow and unsteady motion. To ensure minimal friction we opted for a hanging droplet arrangement: the motion of the particles took place on the bottom surface of a drop of water held by surface tension. The curvature of the hanging drop had no observable effect on the motion of the sailboats.

The driving light was provided by a near infrared laser (IPG YLS-1070, I= 10 W, $\lambda$= 1070 nm). The experiments were carried out in a Zeiss Axio Observer A1 inverted microscope. The laser light reached the sample from above, through the condenser. The diameter of the illuminated area could be easily varied, in most experiments it was about 300 µm, with either Gaussian or top hat intensity profile.

The light driven motion of microstructures in the water sample was observed in a microscope and recorded on video. When turning on the illuminating laser, the objects started to move. To characterize the motion of individual particles, a low density of particles was used and the movement of swimmers was observed in the condition where interaction between different particles was sufficiently rare to allow for efficient data collection and analysis of independent motion. Within the recorded video data we indentified time and spatial regions where objects moved freely for >5 s (typically >15 s) without colliding with each other. A large number of observations were performed to establish a statistical characterization of the mechanical behavior as well. Fig.3. shows a collection of particle motion trajectories. Each individual trajectory was measured in the same sample, however, independently. At the beginning, the laser being turned off, the sample was moved to position the observed particle correctly with respect to the illuminated area. Then the driving laser was turned on, the motion was initiated and recorded. With this procedure all trajectories were measured under identical illumination conditions, and e.g. the data shown in Fig.3. could be compared directly.

The major characteristic properties of the motion can be determined immediately. The particles move randomly, there is no dominant direction. The motion has a persistent character, in takes place in a steady, more or less well determined direction. Consequently, they behave as independent swimmers. By observing the velocity distribution in the illumination area it is obvious that the speed is proportional to the illumination intensity: they move faster when the intensity is larger. To characterize the motion it is also necessary to determine how the direction of motion corresponds to the position of the object, how it is determined by the orientation of the reflective surface. This was achieved by determining the orientation of the elongated particles and it was compared with the direction of motion. Fig. 4. shows the angle between the long axis of the structure and the tangent to the trajectory at any point of the motion as seen in the movie recordings. The good correlation is consistent with the basic interpretation of the observations, i.e. that the driving force is provided by the reflection of the light coming from above.

We can estimate the efficiency of the system in terms of momentum transfer between the incoming light and the moving sailboats. The 45 degree geometry of the objects provides optimal momentum transfer for the reflection of light coming vertically. However, looking from above only part of the visible surface of the sailboat will reflect the light into the proper direction. Nevertheless, we can define an overall efficiency that relates the momentum of light hitting the total illuminated area to the resultant velocity of the sailboat. In the experiments shown in the figures a light intensity of 1.5 W reached the sample in a circular area of about 100 μm diameter, this gives an input pressure of 0.6 N/m$^2$ in this region. The surface of a single sailboat from which the pushing light is reflected looking from above is 45 μm$^2$, thus the force to be converted to move the sailboat is 27 pN. We approximated the resistance of motion by the Stokes formula: $F = 6\pi\eta r v$, (we neglected the friction on the water-air surface). In the Stokes formula we used an effective radius of 5 μm giving the actual frontal area of 80 μm$^2$ of the object. The above numbers yielded a velocity of 280 μm/s, i.e. an ideal system would exhibit this number. Our particles moved somewhat slower (typically 10 to 100 μm/s), but in the calculated order of magnitude (see Fig.3.). Given the approximate nature of the above estimation – note that in the calculation we did not even take into account the momentum of the light reflected in the direction opposite to the propagation, as depicted in Fig.1. - and the uncertainty in the parameters (primarily the exact value of the actual laser power density and the neglecting of friction), we see that a reasonable momentum transfer was observed, i.e., the system is quite efficient.

Additional important characteristics of the motion is regularity. The particles move substantially in steady directions determined by their orientations, however, they tend to change their directions with time due to several possible factors. Rotational diffusion as a result of fluctuating torques exerted on the particle by the medium and imperfections in the microstructures and illumination can all contribute to the observed variation of direction. To characterize this feature, we calculated the persistence length P for the measured trajectories using the abstract definition of P, being the length over which correlations in the direction of the tangent are lost. Using the expression

$< \cos(\theta) > = \exp( -L/P )$,

where θ is the angle between two points of the trajectory separated by a contour length of L, we obtained values for P in the range of 150 μm to 1500 μm for the different trajectories. We also applied the Ornstein-Uhlenbeck model (21), describing the motion of a massive Brownian particle under the influence of friction, to quantify the persistence length. According to Fürth's formula, the mean square displacement of the particles after an elapsed time t can be expressed in terms of the diffusion coefficient D and the persistence time $T_p$:

$$<\mathbf{d}(t)^2> = 4 D [ t - T_p ( 1 - \exp(-t/T_p) ) ]$$

For short and long time periods, the formula simplifies to:

$$<\mathbf{d}(t)^2> = 2 D / T_p * t^2 \qquad , t << T_p$$

$$<\mathbf{d}(t)^2> = 4 D t \qquad , t >> T_p$$

The experimental *d(t)* curves can be associated with a persistent random walk behavior described by the above expressions: the microscopic sailboats move at an approximately constant speed of sqrt( $2D / T_p$ ) for short time periods, while the motion is diffusive when investigated over a long time. By fitting Fürth's formula to all experimental trajectories of a characteristic experiment, we obtained a value of 40.66±0.47 sec for the persistence time corresponding to a persistence length of 355.09 μm (converted using the average speed of the particles).

The variation of the parameters has to be commented: ideally one might expect regular behavior with statistical perturbations, i.e. rather steady – straight - motion characterized by very large persistence length. In our experiments the directions changed in a rather well defined, systematic looking manner, with shapes characteristic for individual trajectories. We believe that the reason is that the motion is very sensitive for irregularities in the sliding surface, very small contaminations have a profound and systematic effect on the motions. Consequently, the motion can not really be regarded as a random walk. Still, we analyzed the motion in such terms, to be able to make a quantitative comparison with other systems. In our system the motion was quite steady so that it could be regarded as highly regular, sufficient for further experiments.

The presented system is an addition to several ones introduced previously. The significantly distinct characteristics offer different applications. Comparison of various designs should be made on very different bases. It may be interesting to compare the propulsion efficiency of the swimmers. A useful practical efficiency parameter is the velocity of the motion generated by a given illumination light intensity. In the case of our sailboats a light intensity of $10^8$ W/m$^2$ generated a motion with about 10 μm/s velocity. As an example, in the case of the Janus-particles driven by thermophoresis due to laser illumination (10) a velocity of about 1 μm/s was driven by a light intensity of about $10^8$ W/m$^2$. The propulsion efficiency as defined above is thus one order of magnitude higher in our case. In the other referenced alternative systems it is not so straightforward to make a reliable estimation of this kind.

We conclude that we have created a system of swimmers where individual particles move in an autonomous manner. The energy of motion is supplied by a radiation field, laser light illumination from above. The particles use the energy flow in a way that is independent of the radiation direction, they move in a direction only determined by their own orientation. Based on macroscopic analogy we call the moving particles "sailboats". The system is very efficient, exceeding the comparable ones introduced recently. Our particles move with a very large persistence length, this may be advantageous in a number of applications. We believe this system can be effectively used to study the motion of self propelled particles, either individual or groups. It may be especially suited to study the collective motion of large number of moving objects.

**Acknowledgements**

This work was supported by the grants OTKA K 84335 (to AB, LK, AM, LO, GV, PO) and EU ERC COLLMOT-227878 (to TV). We thank E. Mihalik for the SEM images of the microstructures.

**Figure captions**

FIG 1. **Experimental arrangement and function scheme of the light driven microrobots**. **a.** Overall view of the microstructures illuminated by a collimated vertical downward directed light beam. The objects are driven by the momentum change of the reflected light in directions determined by their position. **b.** The details of the momentum change of the illuminating light upon reflection. Note that part of the light is reflected in a direction opposite to the direction of motion.

FIG 2. **Light sailboats.** (a) Geometry of the photopolymerized microstructures, (b) electron microscope image of the wedges on a glass surface after the photopolymerization process, (c) closeup of a wedge in the position of action, i.e. resting on one of its planar sides, (d) the final sample: sailboats under the microscope.

FIG 3. **Light sailboats in motion.** Position, velocity and orientation as determined by video analysis. (a) Typical trajectories in the illumination area (selection of 15 objects). (b) Closeup of a trajectory showing the fluctuating orientation of a sailboat while sliding. The circles represent the centers of the microstructure at equidistant points in time (1 second), the line shows the long axis of the elongated object, the ellipse visualizes the actual size of the object, the arrow indicates the direction of motion.

FIG.4. **Position of the sailboats while moving.** Correlation of the direction of motion with the orientation of the saliers. Histogram of the angle between the long axis of the light sailers and the direction of motion for the data shown in Fig.3.

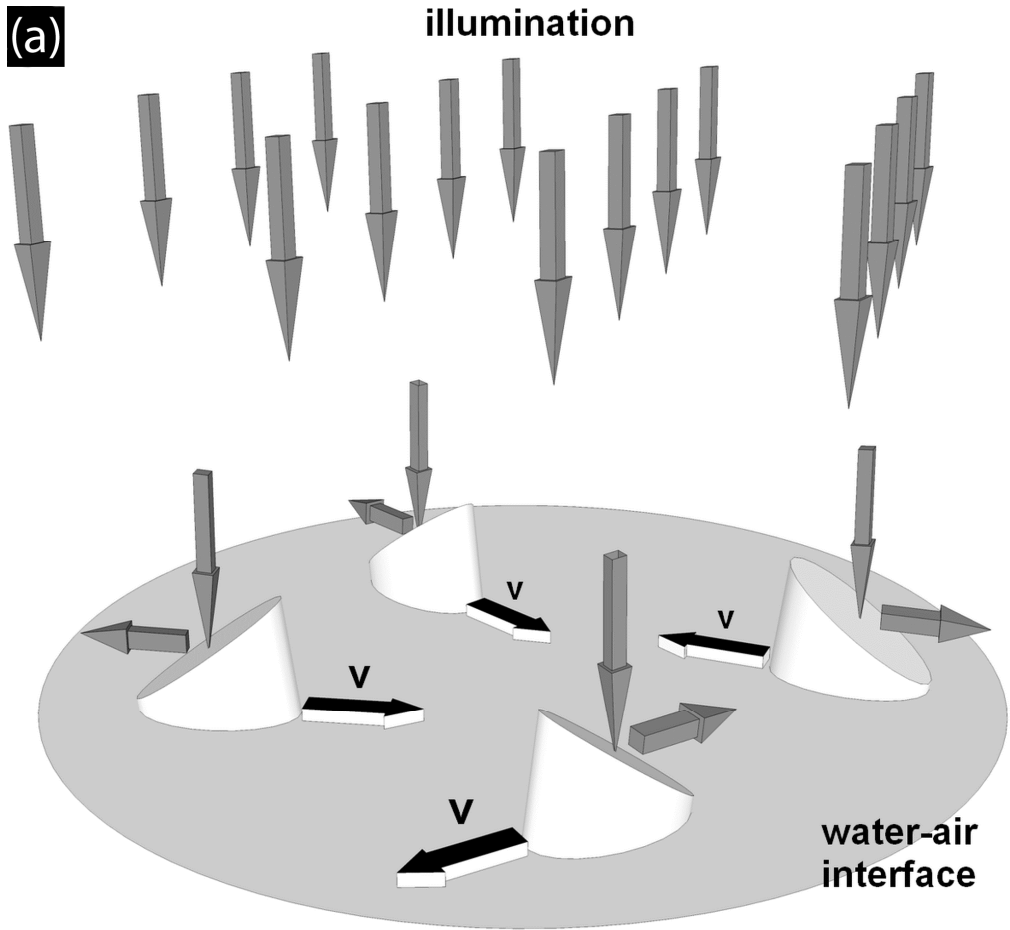

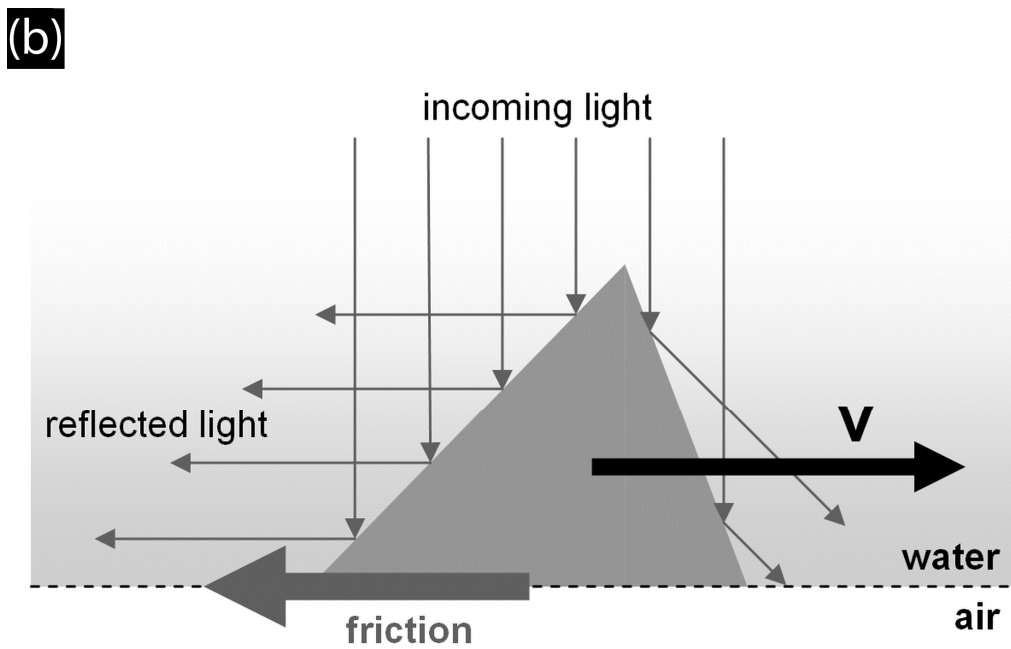

Fig.1.

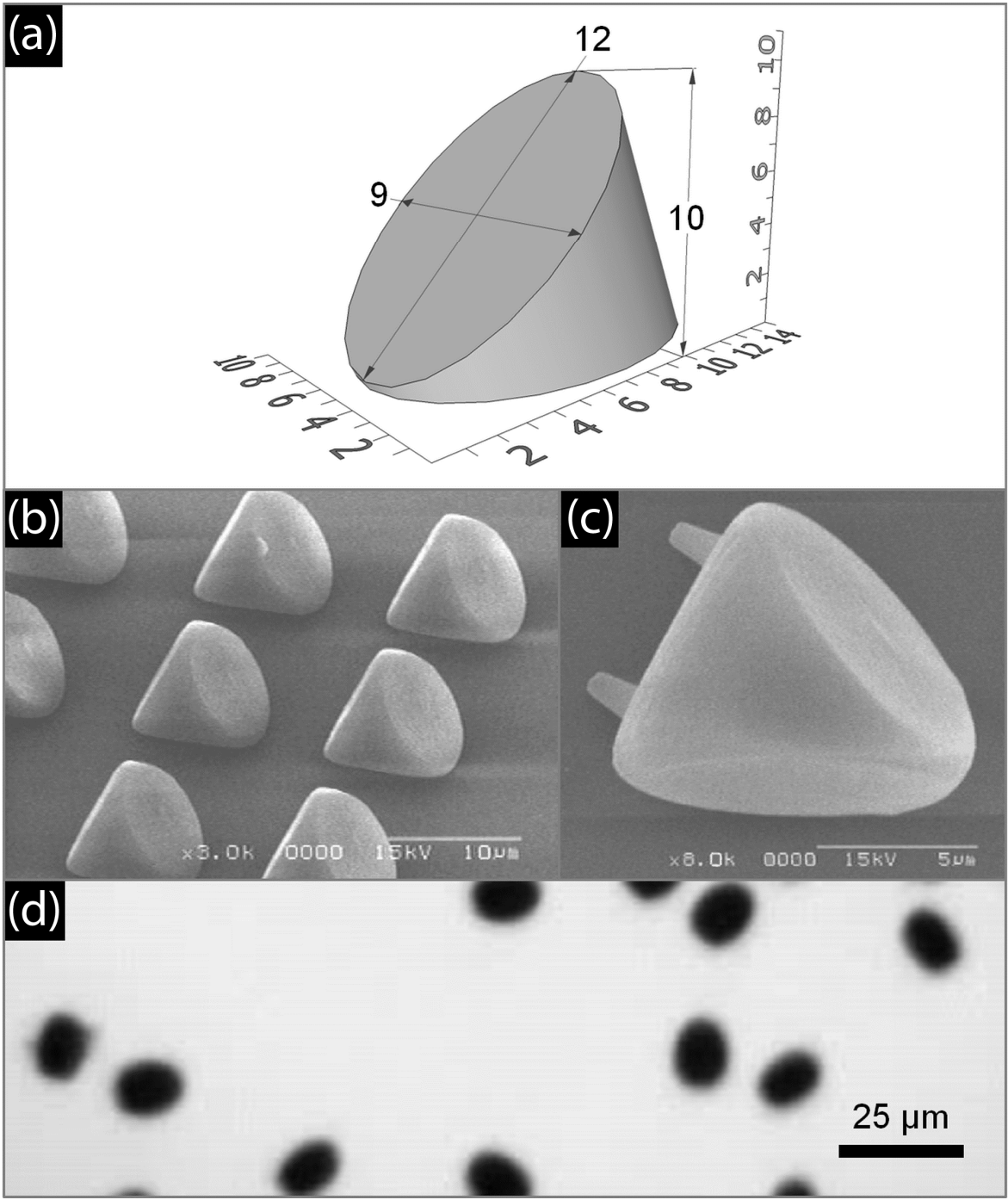

Fig.2.

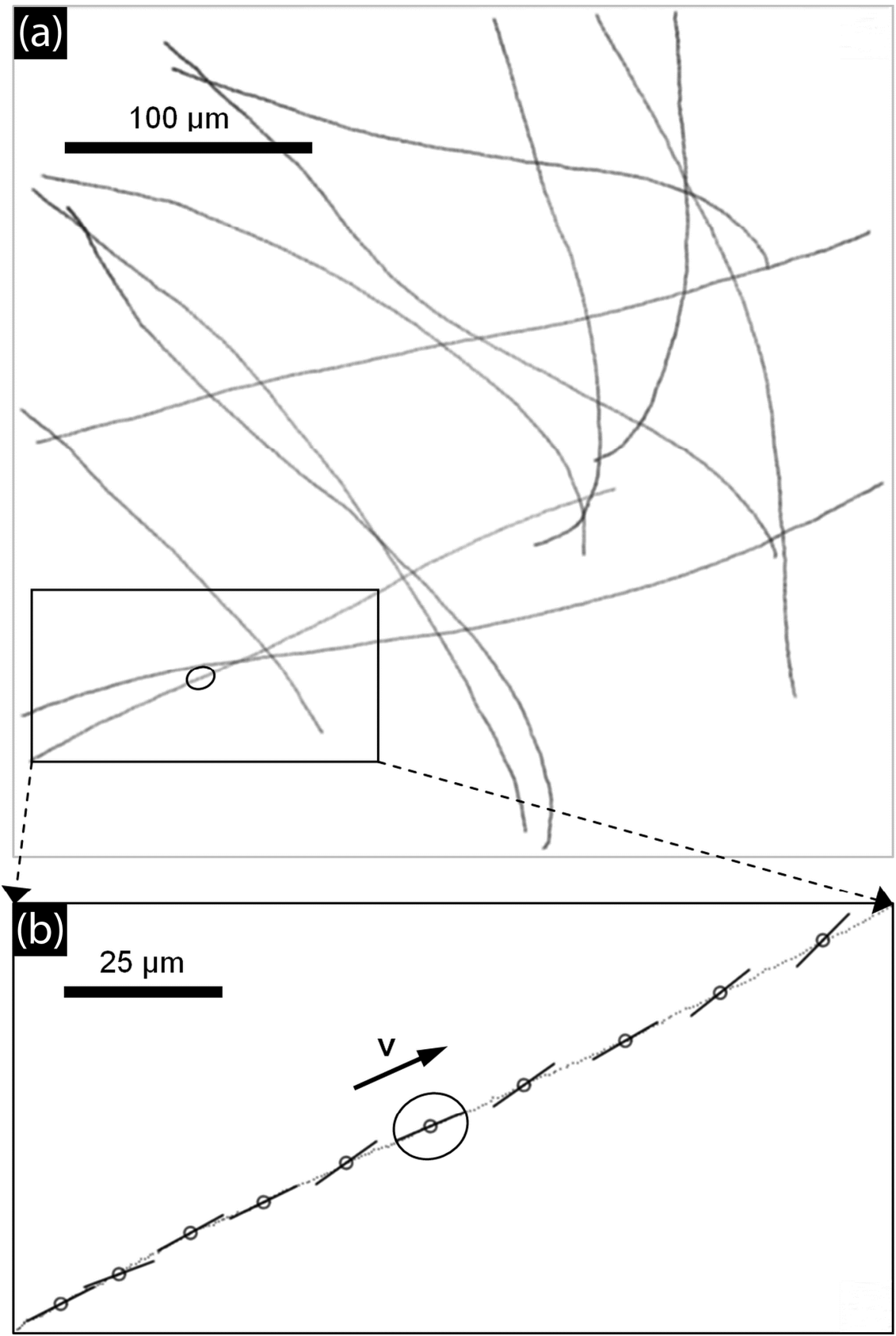

Fig. 3.

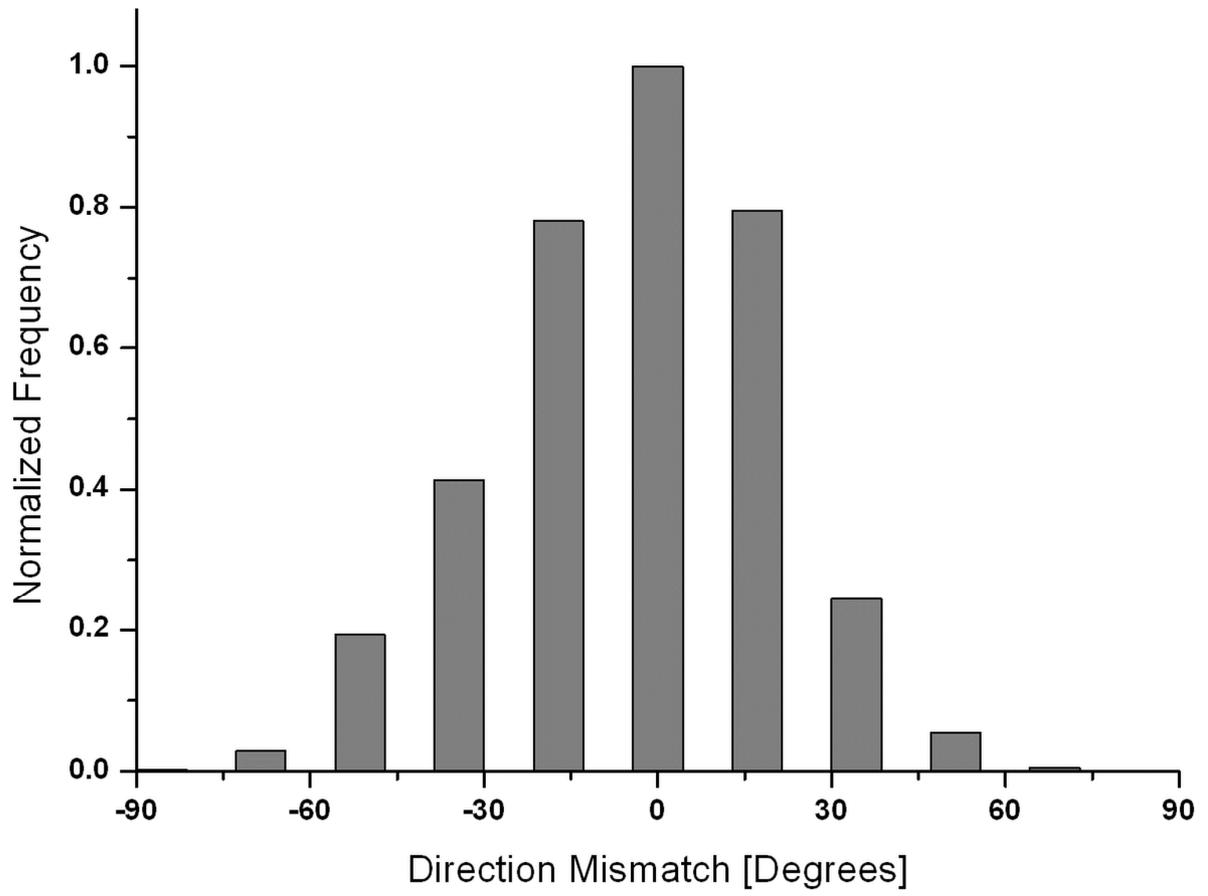

Fig.4.